\documentclass[aps,pre,twocolumn,amsmath,amssymb,showpacs,showkeys,floatfix]{revtex4-1}
\usepackage{epsfig}
\usepackage{graphicx}
\usepackage{bm}
\usepackage{amsmath}
\usepackage{physics}
\usepackage{xcolor}
\begin{document} % INITIALIZE - DONT CHANGE

\title{Control of dynamical localization in atom-optics kicked rotor}
\affiliation{Department of Physics, Indian Institute of Science Education and Research, Pune 411008, Maharashtra, India}
\author{S. Sagar Maurya}
\affiliation{Department of Physics, Indian Institute of Science Education and Research, Pune 411008, Maharashtra, India}
\author{J. Bharathi Kannan}
\affiliation{Department of Physics, Indian Institute of Science Education and Research, Pune 411008, Maharashtra, India}
\author{Kushal Patel}
\affiliation{Department of Physics, Indian Institute of Science Education and Research, Pune 411008, Maharashtra, India}
\author{Pranab Dutta}
\affiliation{Department of Physics, Indian Institute of Science Education and Research, Pune 411008, Maharashtra, India}
\author{Korak Biswas}
\affiliation{Department of Physics, Indian Institute of Science Education and Research, Pune 411008, Maharashtra, India}
\author{Jay Mangaonkar}
\affiliation{Department of Physics, Indian Institute of Science Education and Research, Pune 411008, Maharashtra, India}
\author{M. S. Santhanam}
\email{santh@iiserpune.ac.in}
\affiliation{Department of Physics, Indian Institute of Science Education and Research, Pune 411008, Maharashtra, India}
\email {Electronic mail: santh@iiserpune.ac.in}
\author{Umakant D. Rapol}
\email{umakant.rapol@iiserpune.ac.in}
\affiliation{Department of Physics, Indian Institute of Science Education and Research, Pune 411008, Maharashtra, India}
\date{\today}

\pacs{physics}
\begin{abstract}
Atom-optics kicked rotor represents an experimentally realizable version of the paradigmatic
quantum kicked rotor system. After a short initial diffusive phase the cloud settles down to a stationary state due to the onset of dynamical localization. In this work we realise an enhancement of localization by modification of the kick sequence. We experimentally implement the modification to this system in which the sign of the kick sequence is flipped by allowing for a free evolution of the wavepackets for half the Talbot time after every $M$ kicks. Depending on the value of $M$, this modified system displays a combination of enhanced diffusion followed by asymptotic localization. This is explained as resulting from two competing processes -- localization induced by standard kicked rotor type kicks, and diffusion induced by  half Talbot time evolution. The evolving states display a localized but non-exponential wave function profiles. This provides another route to quantum control in kicked rotor class of systems. The numerical simulations agree well with the experimental results.
\end{abstract}

\maketitle

%\section{Introduction}
The kicked rotor (KR) has been extensively investigated as a paradigmatic model of both classical and quantum chaos \cite{CasChi95}. The atom-optic kicked rotor (AOKR) is an experimentally realisable analog of the KR model in which an ensemble of cold atoms are periodically kicked by sinusoidal potentials formed by a standing wave of light \cite{MooRobBha95, AmmGraShv98, RinSzrGar00, ArcGodObe01}. The classical limit of AOKR is chaotic for sufficiently strong kick strengths and exhibits intrinsic stochasticity and  diffusive growth of mean energy. In contrast, the quantum regime entirely suppresses the classical diffusive growth beyond a short break-time due to destructive quantum interferences analogous to Anderson localization in real space. This shows up as localization of wavefunctions  in  momentum space, $\psi_p \sim e^{-p/\xi}$, where $p$ labels discrete momentum states and $\xi$ is the localization length. Further, AOKR and its variants are studied in other fields -- condensed matter physics \cite{FisShmGre82,BenCasGua01}, molecular physics \cite{BluFisSmi86,AveArv01,ShaBru00,TotCse2021},  quantum information \cite{FacPasSca99,GeoShe01}, exploration of quantum correlations and  many-body effects.

In the context of AOKR, the ability to control localization length and the saturated energy of the localized states can be useful \cite{BitMil17} and has direct implications in the emerging area of quantum technologies. In the context of dynamical localization it has been shown that the addition of stationary noise to system parameters or spontaneous emission \cite{OttAnt84,Coh91,GraMiy96,KlaOskSte1998,AmmGraShv98,SteMilOsk2000,MilSteOsk2000,OskSteRai2003}, coupling with rotor \cite{SanArn20}, and quantum measurements \cite{FacPasSca99} can destroy the delicate nature of dynamical localisation. However, surprisingly, it was shown experimentally that L{\'{e}}vy noise added to kick sequences of AOKR could control the decoherence rate and even the mean energy of localization, the L{\'{e}}vy parameter in the noise distribution acting as the control parameter \cite{SarPauVis17}. More conventional routes to exercise control is by manipulating the phases of the initial wavefunctions, as shown in refs. \cite{GonBru01(2), GonBru01}   for suppressing or enhancing  quantum-chaotic behaviour. Control of localization requires  the unitary evolution operator, that evolves the initial state to change, which is not easily achievable in experiments by just controlling the phases. In one such control scheme, introduced by Gong and Brumer \cite{GonHanBru03}, the phase of the kicking field is flipped periodically.
This variant of AOKR is different from the amplitude-modulated kicked rotors in which the kick strength is a stochastic variable drawn from a suitable distribution \cite{KlaOskSte1998,SteMilOsk2000,MilSteOsk2000}. 

In this work, we experimentally realize a protocol of quantum control (similar in spirit to the one introduced in Ref. \cite{GonHanBru03}) of diffusive and localized phases by appropriate modulation 
of the perturbations. Effectively, the sign of the kick strength in the kicked rotor system is periodically flipped. This is achieved using periodic, time delayed kicks after a certain number of standard kicks that induce dynamical localization. Quite remarkably, this simple modification of kick sequences in AOKR does not destroy localization but leads to an enhancement of quantum energy at which localization takes place. In contrast to the earlier work \cite{GonHanBru03} that depends on presence of classical transporting islands in phase space for energy enhancements, in this work
we show that enhancements arise from a competition between two periodic kick sequences -- one that supports localization and the other that supports diffusion -- in the AOKR system. 

\begin{figure}[t]
\includegraphics*[width=1\linewidth]{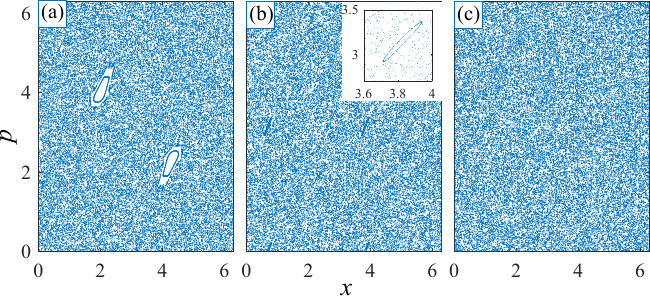}
\caption{Poincar\'{e} sections of the standard ($M=0)$ and modified kicked rotor for kick strength $K=5$ and (a) $M=0$, (b) $M=2$ and (c) $M=3$. The regular islands in the chaotic sea in (a) are non-transporting (b) transporting in nature, albeit much smaller in size. Inset in (b) shows an enlarged view of one of these islands. No regular structures are visible in (c).}
\label{fig:1}
\end{figure}

%\section{Modified atom-optics kicked rotor}
The system of interest is a modification of kicked rotor given by  \cite{GonHanBru03},
\begin{eqnarray}
H=\frac{p^2}{2} + K \cos(x) ~ \sum_{n}f_{M}(n) ~ \delta(t-n),
\label{eq1}
\end{eqnarray}
where $p$ and $x$ denote the dimensionless momentum and position respectively, 
$K$ is the chaos parameter, and time $t$ is scaled by the pulse period $T$ such 
that $t \rightarrow t/T$. If  $f_{M}(n)=1$, then Eq. \ref{eq1} is just the standard 
KR. In this work, $|f_{M}(n)|=1$, and $f_{M}(n)$ changes sign after every $M$ kicks. In rest of this paper, Eq. \ref{eq1} will be referred to as the Modified Kicked Rotor (MKR) model,
and it can be thought of as a specific realisation of a generalised KR model in Ref. \cite{DanEisShn96}. The classical stroboscopic map for the phase space variables for MKR \cite{GonHanBru03} is shown in Fig.\ref{fig:1}.

As evident from Fig.\ref{fig:1}(a), the phase space is largely chaotic with a few regular islands. An ensemble of initial conditions launched from these islands will remain bound to these islands. In the case of MKR with $M=2$, special non-chaotic structures called transporting trajectories exist in phase space, as seen in the inset of Fig. \ref{fig:1}(b). They are present in many periodically driven dynamical systems, {\it e.g.}, standard map \cite{Izr90}, Hamiltonian ratchets \cite{SchOttKet01}, atomic KR \cite{SteOskRai01,HenHafBro01} and are usually referred to as the accelerator modes because they support anomalous classical diffusion, {\it i.e.}, $\langle E \rangle_n \propto n^2$.  
 The quantum dynamics of MKR is obtained through the use of split operator technique \cite{Izr90}.

\begin{figure}[t]
\includegraphics*[width=0.6\linewidth]{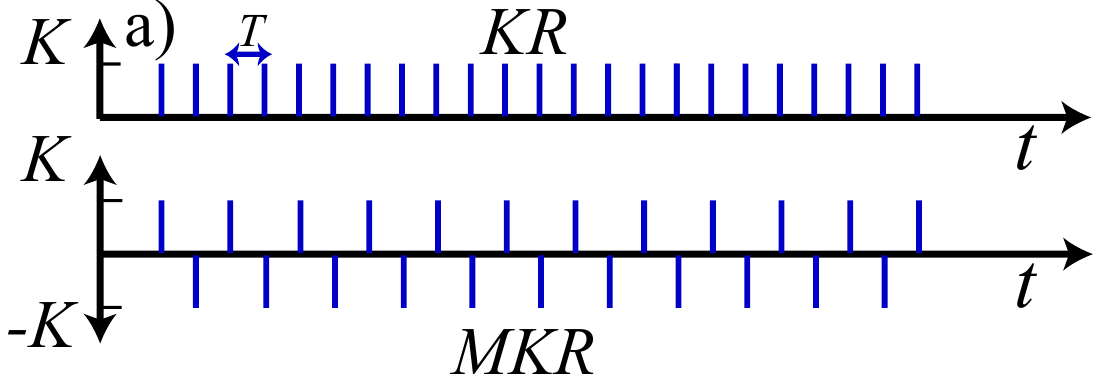}
\includegraphics*[width=0.8\linewidth]{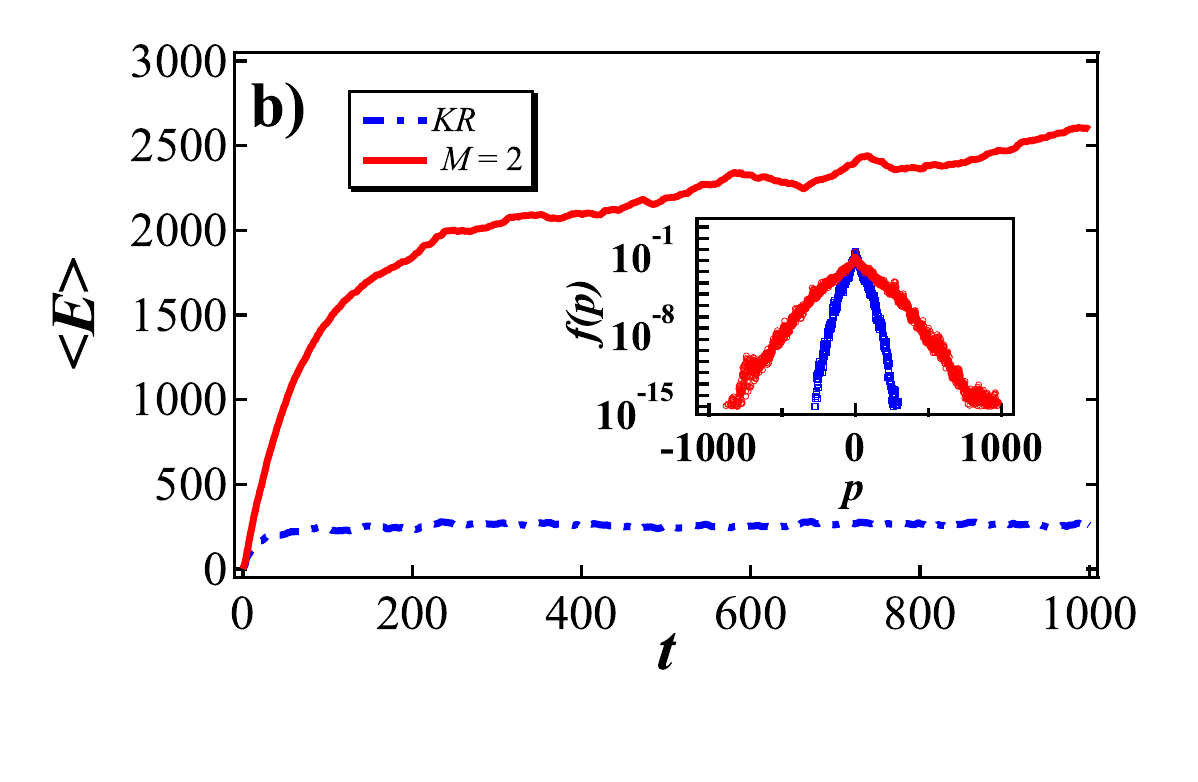}
\caption{ (a) Pulse sequence for the standard and the modified kicked rotor with 
$k=5$ and $\hbar_{\rm eff}=1$ ($K=5$). For $M=2$,  the sign of $K$ is flipped at every kick.
(b) Simulated energy evolution of KR (blue solid line) and MKR with $M=2$ (red solid line). Inset shows the momentum distribution for the same at $t=1000$. Open squares are for KR and open circles are for MKR with $M = 2$. }
\label{fig:2}
\end{figure}

\begin{figure}[t]
\includegraphics*[width=0.6\linewidth]{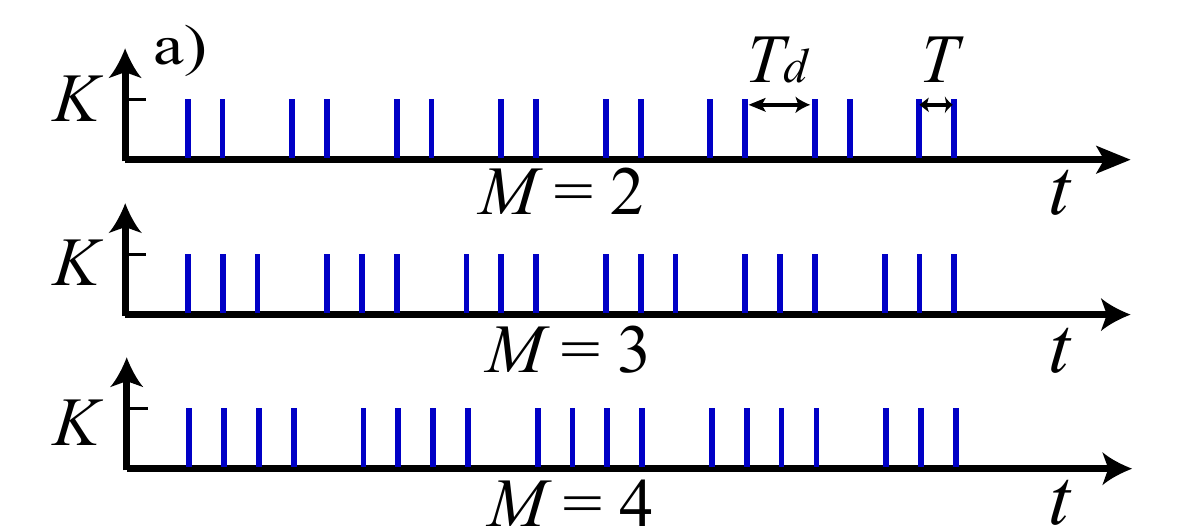}
\includegraphics*[width=0.8\linewidth]{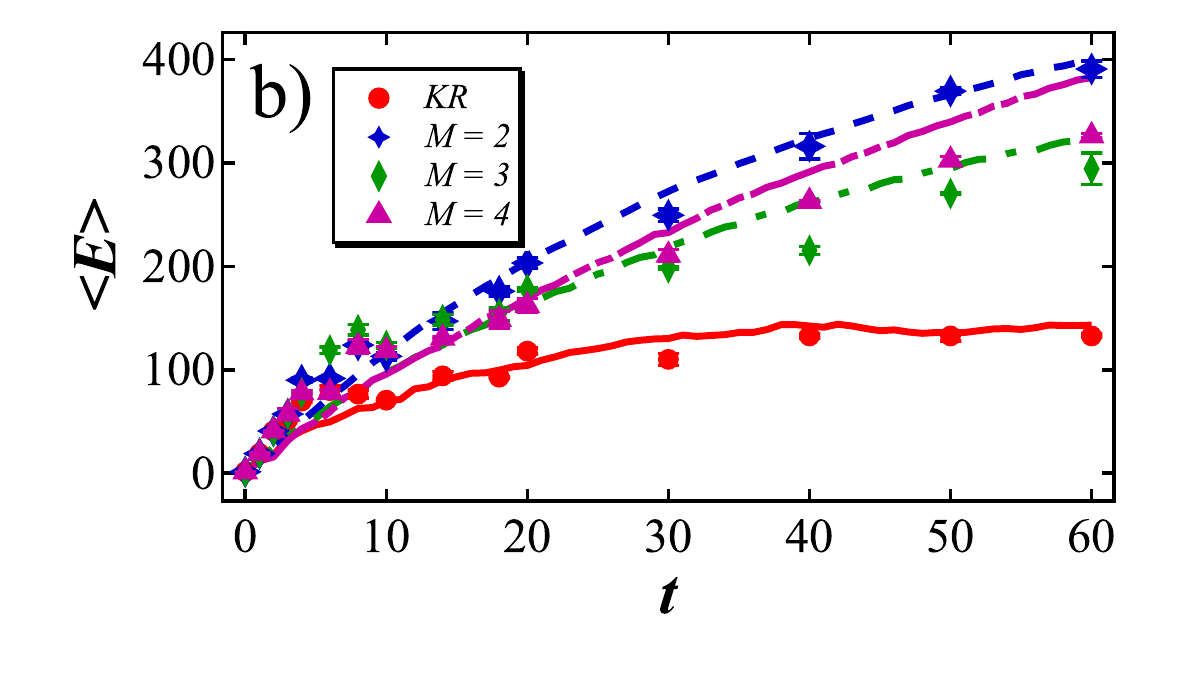}
\caption{a) The kick sequence for different values of M. b) Mean energy vs. $M$ at $k=5$ and $\hbar_{\rm eff}=1$.}
\label{fig:3}
\end{figure}

To build the quantum dynamics of MKR, let us consider the period-1
Floquet operator corresponding to the standard kicked rotor $\widehat{F}_{\rm KR}^{\pm}$ 
for which $f_M(n)$ is a constant, {\it i.e.}, $f_M(n)$ is either $+1$ or $-1$ for all $n$. 
Thus, the required operator is
\begin{equation}
\widehat{F}_{\rm KR}^{\pm}(T) =
\exp\left[i\frac{\hbar_{\rm eff} T}{2}\frac{\partial^{2}}{\partial x^{2}}\right]
\exp[\mp i\frac{K}{\hbar_{\rm eff}}\cos(x)],
\end{equation}
where $\hbar_{\rm eff}=T\hbar$ is the scaled Planck's constant, $k=K/ \hbar_{\rm eff}$ denotes the
strength of phase modulation imparted by the kicks.
Using this as the building block, the Floquet operator for MKR can be constructed as
$M$ application of $\widehat{F}_{\rm KR}^{+}$ followed by $M$ application of $\widehat{F}_{\rm KR}^{+}$. Thus, for MKR, we obtain
\begin{equation}
\widehat{F}_{\rm MKR}(T) = \left[ \widehat{F}_{\rm KR}^{-}(T) \right]^M ~ \left[ \widehat{F}_{\rm KR}^{+}(T)
\right]^M
\label{eq:Fmkr}
\end{equation}
The kicking scheme for implementing the above Floquet Hamiltonian (Eq. \ref{eq:Fmkr}) is shown in Fig. \ref{fig:2}(a) and simulated time 
evolution of mean energy is shown in Fig. \ref{fig:2}(b) for KR and MKR with $M=2$. The inset of 
Fig. \ref{fig:2}(b) shows the momentum distribution.
Since abrupt phase changes are technically challenging,
an alternative method to realize MKR is by introducing controlled time delays 
after every $M$ kicks \cite{GonHanBru03}. Consider a wave-function of the system 
$\Psi(x,t)$ at any time $t$. This can be expanded in the momentum basis as
$\Psi(x,t)= \sum_{m} A_{m} \langle x|m\rangle$ with $A_m$ being the expansion coefficients.
The flipping of sign of $K$ can also be thought of as shift in spatial coordinate 
by $\pi$, since $ K \cos (x+\pi)= -K \cos x$. Hence, it is convenient to use
$ x \to x+\pi$ instead of $K \to -K$. 
Through simple manipulations, it can be shown that the change of sign of kicking strength effectively introduces a 
phase difference of $\pi$ between the neighbouring states. This phase difference can also
be generated by introducing time delays in the system. To see this, consider the
 action of a free-evolution operator of MKR for $t=T$ acting on a momentum state $\exp\left(ip^{2}T/2\hbar \right) ~ |m\rangle = \exp\left( im^{2}\hbar T/2 \right) ~ |m\rangle$.
From this, we can estimate the duration of free evolution $T_d$ (called delay time) required to obtain a phase-difference of $\pi$ between neighbouring momentum states. For a phase
difference of $\pi$, the condition to be satisfied is 
$\exp{i\hbar d[(m+1)^{2}-m^{2}]/2 } = \exp(i\pi)$, and from this we get the
time duration to be $T_d= {2\pi}/{\hbar} = {2\pi T}/{\hbar_{\rm eff}}$.
This delay time $T_d$ corresponds to half Talbot time and has the same effect as 
flipping the sign of the kick strength between the pulses \cite{GonHanBru03}. 
Figure \ref{fig:3}(a) shows the kicking scheme for KR (blue), and MKR with $M=2$ (red),
and MAKR with $M=2$. The kick period $T$ and delay time $T_d$ are also indicated in this figure.
Due to the presence of two periods, the system is no longer periodic with time 
period $T$, but has an effective period of $T(M-1)+T_d$. Thus, the Floquet 
operator  $\widehat{F}_{\rm MAKR}$  for the Modified {\it Atom Optic Kicked Rotor} (MAKR) is given by:
\begin{multline}
\widehat{F}_{\rm MAKR} = \widehat{F}_{\rm KR}^{-}(T_d)  \left[ \widehat{F}_{\rm KR}^{-}(T) \right]^{M-1}
= \widehat{\bf{F}}_{\rm KR}\widehat{F}_{\rm KR}^{M-1}.
\label{floq_MKR}
\end{multline}
In this, $\widehat{F}^{M-1}_{\rm KR}$ denotes $M-1$ applications of $\widehat{F}_{KR}$ 
for time duration $T$, and $\hat{\bf{F}}_{KR}$ denotes the application of kicked rotor Floquet operator with a free propagation time $T_d$. 

%\section{Atom-optics experimental setup}
The variables and parameters of kicked rotor system and AOKR are related as: 
$x \rightarrow 2k_L x$, $p \rightarrow 2k_L Tp/m$, and $\hbar_{\rm eff} = 8 \omega_r T$ 
where $k_L, m$ and $\omega_r$ are the wavenumber of the lattice laser, mass of the atoms and recoil frequency respectively. The kick 
strength is $k = \hbar \Omega^2 \tau / 8 \Delta$, where $\Omega$, $\Delta$ and $\tau$ 
are the resonant Rabi frequency, the detuning of the light used to create the optical 
lattice potential and the pulse duration respectively. 
The experimental setup and the implementation sequence is the same as in Ref. \cite{SarPauVis17}. 
To synthesize the MKR Hamiltonian, the off-time between the pulses is adjusted. The on-time $\tau$ of the standing wave is kept as $100$ ns. 

To keep $K$ fixed, we adjust $k$ for different $\hbar_{\rm eff}$. To calibrate the kick strength $k$, we use Raman-Nath diffraction on an almost ideal zero-momentum state {\it i.e.}, 
a Bose-Einstein Condensate. By fitting the number of atoms in the $n^{\rm th}$ momentum state after undergoing diffraction is a Bessel function $J_{n}^{2}(k)$, $k$ can be accurately 
determined \cite{JayCheSag20}. Numerical simulations using Eq. \ref{floq_MKR} are performed on a thermal cloud with gaussian momentum distribution.

%\section{Evolution of mean energy}
\label{expt_th_MKR}

%\subsection{$M=2$ case}

{\bf Case of  $M=2$ : }The time evolution of mean quantum energy from the numerical simulations
of MKR using Eq. \ref{floq_MKR} is shown in Fig. \ref{fig:2}(b).
For the standard KR, the mean energy displays the expected linear increase for time periods less than the break time $t_b$. For $t > t_b$, quantum effects lead to dynamical localization. For the same set of parameters as
that for KR, except that for $M=2$, a pronounced enhancement in the saturated energy 
is evident in this figure. Dynamical localization occurs for $M=2$ as well.
In case of MKR, with $M=2$ and $K=5$, transporting trajectories are present in phase space
(Fig. \ref{fig:1}(b)). Thus, as argued in Ref. \cite{GonHanBru03}, in the quantum regime, the mean energy corresponding to MKR is enhanced with respect to the standard KR. Further, the width of momentum distribution (inset in Fig. \ref{fig:2}(b,c)) 
is also significantly enhanced for MKR. 

\begin{figure}[t]     
\includegraphics*[width=0.8\linewidth]{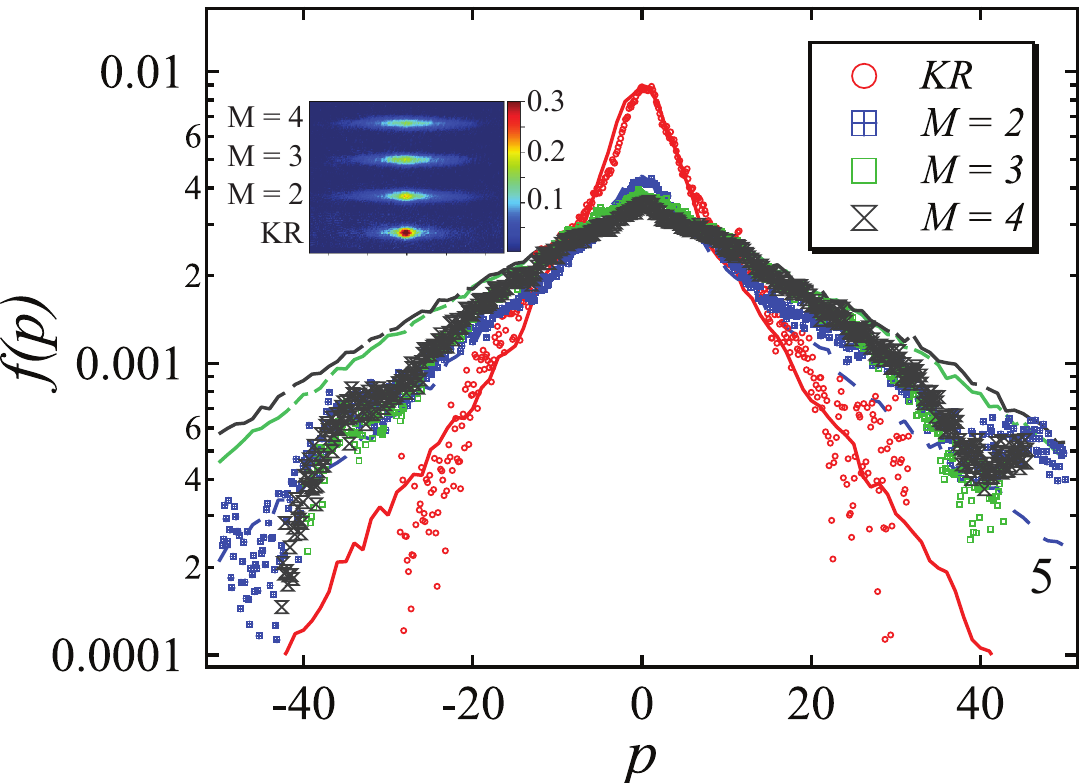}
\caption{Momentum distribution profiles for $M=0$ and 2 at the parameters $K=5$ and $\hbar_{\rm eff}=1$. The dots and the solid lines denote experimental and simulated momentum profiles respectively. Profiles for $M=2, 3$ and $4$ clearly show enhanced localization length over that of $KR$. Inset shows absorption images (optical density of atoms) for different values 
of $M$ for $\hbar_{\rm eff}=1 $ and $K=5$.}
\label{fig:4}
\end{figure}

Figure \ref{fig:3}(b) displays both the experimental data (solid symbols) as well as the 
numerical simulations (lines) of MAKR obtained using Eq. \ref{floq_MKR}. A good agreement is seen between the experimental and numerical simulations. At first sight, it is tempting to attribute the energy enhancement seen in the experiment to entirely the presence of transporting islands. However, these islands exist in phase space only over a small range of $K$. In the experiment,
the value of kick strength $K$ suffers from approximately 10\% error that washes
out most of the contribution arising from transporting islands.
Enhancement is largely due to the presence of two time scales $T$ and $T_d$
in the MAKR system. The time evolution with period $T$ and kick strength $k$ is arranged such that it induces localization in the system. For an initial state with significant momentum spread ($w \gtrsim 1 $) a pulse period corresponding
to half Talbot time $T_d$ leads to a linear diffusive growth in energy \cite{SauHalCha07}. The dynamics of MAKR is governed by the competition between these contrasting behaviours of localization and diffusion. 
For a finite number of total kicks $N$ applied to the atomic cloud, the number of (diffusion inducing) free evolution phase with time period $T_d$ is $N_d = N/M - 1$. Number $N_l$ of localization-inducing evolution phases with time period $T$ is $N_l=(M-1)N_d$. Thus, if $M \gg 1$, then $N_l \gg N_d$. In this scenario,
localization effects dominate and diffusion is suppressed (The standard KR limit denoted by the red curve in Fig. \ref{fig:3}(b)). In the other limit,
as $M \to 1$, diffusive growth of energy is strongly favoured over localization.
The competition between these processes determines the enhancement of saturated
energy in MAKR system. In a short time scale, localization 
is destroyed by anti-resonance, which eventually sets into the system 
due to destructive interference in  momentum space.
In particular, the contribution of classical transporting
islands is not very significant. For $M=1$, diffusion is dominant and localization
is completely suppressed. For $M=2$, we get $N_d = N_l$. Hence, we can anticipate 
localization as well as diffusive phase. Consistent with this constraint, 
both the experiment and numerics (blue color in Fig. \ref{fig:3}b) show an enhancement 
induced by the diffusive phase as well as localization in the form of saturated 
mean energy. This argument also implies that 
the enhanced saturated energy should be seen for $M=3, 4$ as well, 
even though classically no transporting islands are present in phase space
(see Fig. \ref{fig:1}(c)). We shall consider these cases now.

%\subsection{case of $M>2$}
{\bf Case of $M>2$:} For $M>2$, $N_l > N_d$. This guarantees that localization can be seen for all $M>2$.
However, since $N_d$ decays as $M$ increases, the diffusive phase weakens.
Hence, at any given number of kicks, the highest energy reached for $M>2$ will
always be lesser than that for $M=2$. 
Fig. \ref{fig:3} displays quantum $\langle E \rangle$ against the number of kicks for 
$K=5$ and $\hbar_{\rm eff}=1$ for MAKR for $M=3$ and $M=4$. 
Even for $M=3,4$, a significant enhancement in the energy is seen. This is purely attributed to the half Talbot time evolutions $T_d$ in the kicking sequence, as the classical phase space is completely chaotic. As argued in the previous section, two competing effects are 
at play -- localization induced by the evolution over time period $T$ and 
diffusion due to time delay $T_d$. The long-time behaviour of 
MKR would exhibit complete localization \cite{GonHanBru03}, even though it is not
apparent in Fig. \ref{fig:3} due to the small number ($\sim 60$) of kicks in the experiment. 

From  Fig. \ref{fig:4}, it is clear that the width of the momentum
distribution is larger for $M=2, 3$ and $4$ as compared to that of standard KR. 
This is also visible in the absorption images that carry more weight in the
higher momentum states. 

\begin{figure}[t]
\includegraphics[width=\linewidth]{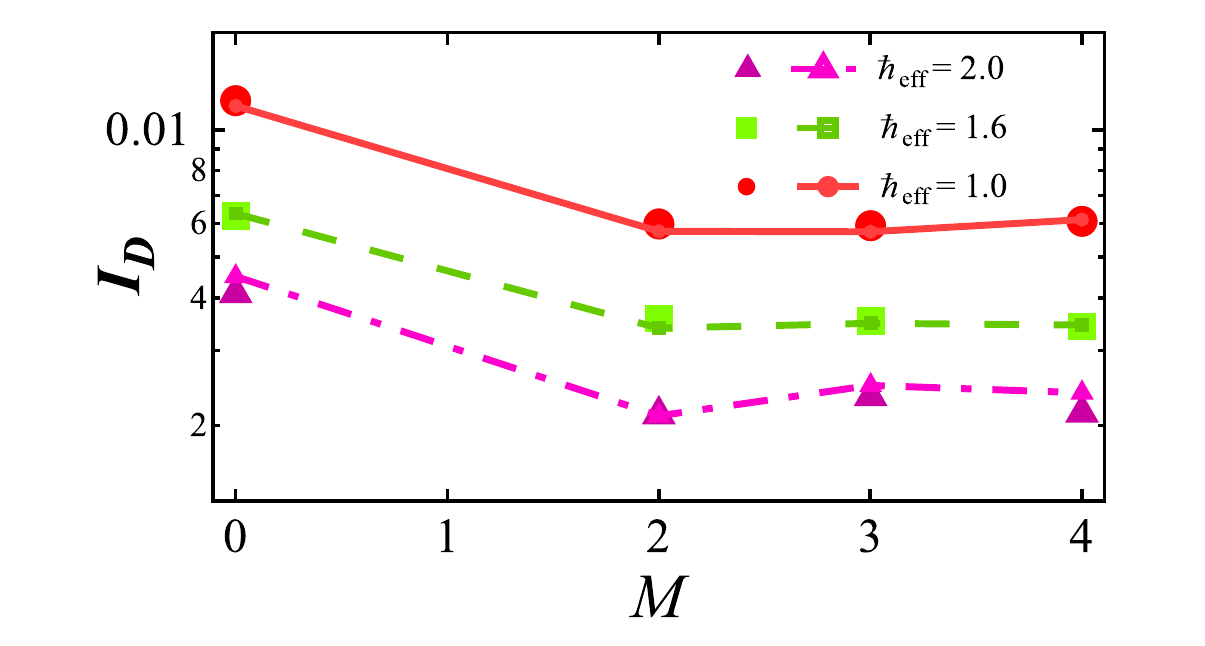}
\caption{IPR for experimental (symbols) and simulated (symbols + dash) data as a function of $M$ for different values of $\hbar_{\rm eff}$  for  a constant  value of $K=5$.}
\label{fig:5}
\end{figure}

For a wavefunction (in momentum representation) $ \langle m | \psi\rangle$, one of 
the commonly used localization measures is the 
inverse participation ratio (IPR) -- $I_D = \sum_{m=1}^D |\langle m | \psi\rangle |^4$,
where $D$ is the dimension of the Hilbert space
in which $\langle m | \psi\rangle$ resides.
A localized wavefunction will have $I_D \sim 1$, while for a completely 
delocalized wavefunction $I_D \propto 1/D$.  

As evident in Fig. \ref{fig:5}, the IPR
for $M=2,3,4$ have a much lower value compared to the KR, implying the spread 
in the momentum distribution and energy enhancement compared to KR. But $I_D$ for 
$M=2,3,4$ are very close to each other, indicating that the spread is almost 
the same for all of them. As transporting islands of significant size are only present 
in the case of MKR with $M=2$ ($k=5$), one would expect to have a maximum energy 
enhancement and momentum distribution spread for $M=2$ \cite{GonHanBru03}. 
But in our case, we observe the momentum distribution to be very close to each other 
for all the $M=2,3,4$. This implies that the enhancement in the localization length or momentum distribution spread arising from transporting islands for $M=2$ is less significant 
in MAKR. Indeed, in this experiment, the bound in the fluctuations in the parameters are sufficiently large enough that in the corresponding classical phase space transporting islands do not exist for $M=2$.
The dynamics is largely determined through the interplay of two time 
periods in the system. It has been shown that very small changes to the system parameters 
($k$ and $\hbar_{\rm eff}$) can lead to the destruction of transporting islands present 
in the classical phase space and non-exponential shape of quantum momentum distribution \cite{GonHanBru03}. This change in the line shape for dynamical localization even occurs 
without causing an obvious difference in energy absorption behaviour. A similar behaviour is observed for various values of $M$ and for a range of parameters $k$ and $\hbar_{\rm eff}$ in the MKR model. The only requirement being that $k$ should be sufficiently large for
dynamical localization to take place in the system. 
As in the standard KR, the experimental data in
Fig. \ref{fig:5} shows that for a fixed $M$, a decrease in $\hbar_{\rm eff}$ is associated with 
broadening of the wavefunction profile. Hence, $I_D$ increases as $\hbar_{\rm eff} \to 0$.

%\section{Conclusions}
\label{res}

In this work, we have investigated a modified kicked rotor model both experimentally and theoretically. The modified atomic kicked rotor is realized by introducing  half Talbot time delays in the kicking sequence, which is equivalent to flipping the sign of the kick strength. It is shown that the modified atom-optics kicked rotor system with $M=2,3,4$ shows enhanced quantum mean energies compared to the standard kicked rotor model. These results  are not only of intrinsic interest in the quantum chaos of kicked rotor, but also in the broader context of quantum control of coherent phenomenon.

\begin{acknowledgments}
UDR would like to thank the Department of Science and Technology, Govt. of India for funding support through the National Mission on Interdisciplinary Cyber-Physical Systems (NM-ICPS). SSM and PD would like to thank the Council of Scientific and Industrial Research, Govt. of India for Research Fellowship. MSS would like to acknowledge funding support from MATRICS grants of the Science and Engineering Research Board of Department of Science and Technology, Govt of India.
\end{acknowledgments}

\end{document}